# Design, construction and evaluation of emotional multimodal pathological speech database

ZHU Ting [1], DUAN Shufei [1*], LIANG Huizhi [2], ZHANG Wei [3]

[1] Taiyuan University of Technology, China; [2] Newcastle University, UK; [3] Taiyuan Hospital, China

zt_april@163.com, duanshufei@tyut.edu.cn, huizhi.liang@newcastle.ac.uk, zwwpa@163.com

## Abstract

The lack of an available emotion pathology database is one of the key obstacles in studying the emotion expression status of patients with dysarthria. The first Chinese multimodal emotional pathological speech database containing multi-perspective information is constructed in this paper. It includes 29 controls and 39 patients with different degrees of motor dysarthria, expressing happy, sad, angry and neutral emotions. All emotional speech was labeled for intelligibility, types and discrete dimensional emotions by developed WeChat mini-program. The subjective analysis justifies from emotion discrimination accuracy, speech intelligibility, valence-arousal spatial distribution, and correlation between SCL-90 and disease severity. The automatic recognition tested on speech and glottal data, with average accuracy of 78% for controls and 60% for patients in audio, while 51% for controls and 38% for patients in glottal data, indicating an influence of the disease on emotional expression.

## 1. Introduction

Dysarthria is a speech disorder due to central neuropathy [1]. Acoustic signals are now widely used in the pathological speech study due to their non-intrusive nature and easy to acquisition [2, 3]. Pathological speech datasets are the basis for the analysis and evaluation of pathological features. The public popular dysarthria speech datasets have Nemours (English, acoustics) [4], AVPD (Arabic, acoustics) [5], SVD (German, acoustics & EGG) [6], and TORGO (English, acoustics & articulation) [7], etc. However, most of them are disease-specific and contain only two modalities. There is a great lack of Chinese pathological speech corpus. Building a Chinese dysarthria dataset can help to study the pathology of dysarthria patients in Chinese regions, which suggests designing and constructing inclusive and multimodal Chinese datasets is right in front of us.

Single-modality pathological speech recognition is still limited, therefore multi-angle and multi-modal pathological symptom analysis will be more objectively and scientifically sound. To research multimodal pathological speech and emotion recognition, many studies have combined audio, glottal, articulatory, and video data [8-10]. Some of them have explored the inclusion of functional near-infrared spectroscopy (fNIRS) in the field of speech disorders [11,12]. These studies demonstrates that the integration of multimodal data enhances research depth and make the pathology and emotion characterization more accurate and three-dimensional. The primary contribution of this paper is to design and build a Chinese multimodal dataset with audio, video, glottal, fNIRS, and multi-perspective information from Symptom Checklist-90 scale (SCL-90), Mini-Mental State Examination (MMSE) and Frenchay Dysarthria Assessment (FDA). 3D Electromagnetic articulography (EMA-AG501) can present and evaluate physiological articulatory movements; this modality data was not included in this paper due to clinical invasiveness and immobility of the device's accuracy calibration.

The biggest innovation of this dataset is the inclusion of emotional pronunciation. According to clinical research, most dysarthria patients have different degrees of affective disorders. For instance, stroke or Alzheimer's patients may experience psychological states such as depression, anxiety, or others [13,14]. A number of related studies in the medical field have shown a relationship between acoustic features and cortical auditory evoked potentials under different emotions, so that the severity of the disease can be graded [15,16]. Unfortunately, the impact of emotion has not been taken into account in past research on pathological speech, and therapies influenced by emotions have not yet been studied and put into practice in the rehabilitation of dysarthria patients. The pathogenic causes of dysarthria can be better understood with the aid of affective computing, which can also serve to direct clinical diagnosis and rehabilitation therapy [17]. Therefore, it is imperative and highly vital to develop a multimodal emotional pathological speech dataset.

## 2. THE-POSSD database

This Chinese emotional pathological speech dataset was jointly collected by Taiyuan University of Technology and Taiyuan Central Hospital of Peking University First Hospital. The project ethics had been approved by the Biological & Medical Ethics Review of both affiliations. The dataset was shortly named THE-POSSD, representing Multimodal Emotional Database of Post-Stroke Speech Dysfunction Patients in Taiyuan, jointly collected by Taiyuan University of Technology and Taiyuan Central Hospital. Every participant gave their written consent form before any recording session, and they all consented to participate in this study.

### 2.1. Subject recruitment

This project included both normal controls and patients with motor dysarthria, all of whom spoke Mandarin and were between the ages of 40 and 80. There were no documented literacy issues or cognitive impairments, and the gender ratio remained balanced. According to the study's aim, the patients could articulate and moderately express emotions. The SCL-90, MMSE, and FDA were tested on all participants prior to data collection. In total, 68 speakers (39 post-stroke dysarthria patients and 29 typical controls) were recruited in Taiyuan, China. The severity of the disease was primarily categorized into mild, moderate, and severe. Demographic information of all subjects is described in Table 1.

**Table 1**: *Subjects' information*

| Type | Number | Gender | | Age | | | Education | | | Disease | | Total |
|---|---|---|---|---|---|---|---|---|---|---|---|---|
| | | Male | Female | 50~60 | 60~70 | 70~80 | Primary | Secondary | High | Main Cause | Time (years) | |
| Mild | 16 | 10 | 6 | 1 | 5 | 10 | 10 | 6 | 0 | Cerebral hemorrhage & Cerebral infarction | >10  37% | |
| Moderate | 10 | 7 | 3 | 2 | 4 | 4 | 6 | 4 | 0 | | >5  26% | 39 |
| Severe | 13 | 9 | 4 | 2 | 11 | 0 | 8 | 4 | 1 | | >1  35% | |
| Control | 29 | 11 | 18 | 7 | 13 | 9 | 15 | 10 | 5 | None | | 68 |

### 2.2. Emotion selection and induction

Existing emotion databases, such as CISIA(Chinese Affective Corpus) [18], IEMOCAP(English Affective Corpus) [19], Emo-DB(German Affective Corpus) [20] and DEED((English Affective Corpus)) [21], and others, widely used a set of basic discrete emotions to capture data. Given the constraints of emotional expression in patients with different levels of condition, four typical emotions--happy, sad, angry, and neutral--were chosen since they have been shown to be the most significant emotions for dysarthria patients to successfully communicate [22].

THE-POSSD employed an audio-visual combination of emotional segments as a medium for eliciting feelings, and it was augmented by Stanislavsky's emotional memory approach [20, 21] to assist subjects' self-induction. Selected video clips were excerpts from movies, TV series, and variety shows. The first clip was an emotion-led video, and the other three clips were direct emotion-evoking videos, with four videos for each emotion. Simultaneously, the subjects were told they may repeat a line as many times as they desired till they were satisfied with their own performance.

### 2.3. Speech materials

Based on the experience of speech therapists, the material sources were selected from the "Mandarin Phonetics Course" book written by Prof. Du [23]. It consisted of the following sessions: (a) Non-words: 5 single vowels, 2 nasal vowels, and 4 compound vowels in Mandarin. These vowels could help understand the pronunciation abilities of dysarthria patients by the frequent and basic phonemes. (b) Single words: 22 digits and 5 single words aiming to study acoustics like formant transition between consonants and vowels. (c) Phrases: 7 syllabic phrases. It could observe the control capability of vocal organs and muscles during articulation. (d) Restricted sentences: 36 complete and semantic sentences, 3 paragraphs, and 1 picture. Sentences or paragraphs with rich phonemes would explore the pathology of articulation in patients. Narrative passages would be probably a good indicator of the fluency and intelligibility of patients with dysarthria.

In addition to the inclusion of emotional stimuli, which required subjects to engage in basic emotional expression; this database also incorporated long text passages and required variable-speed expression whenever possible. This not only allowed for a more comprehensive assessment of the patient's difficulty in expressing emotion, but also examined the subjects' ability to express themselves fluently. A picture was also selected and subjects were asked to describe it on their own, aiming to assess the patient's ability to express themselves verbally and organize their thoughts when confronted with visual information. There were 27 emotional corpora (9 sentences each for happy, sad, and angry emotions), 9 neutral emotional corpora, and 29 non-emotional corpora (12 nonwords, 5 words, 7 words, 4 long paragraphs of text, and 1 descriptive picture), resulting in a total of 65 recorded speech. Only 27 emotional sentences and 9 neutral sentences were selected for this study.

### 2.4. Data collection and pre-processing

Recording instruments mainly included: (a) Electroglottograph (EGG-D100): to acquire synchronized acoustic and glottal signals. (b) Functional near-infrared spectroscopy (22-channel fNIRS): a NIRS-EEG compatible cap consisting of 10 transmitters and 8 detectors to acquire changes in oxyhemoglobin and deoxyhemoglobin under emotional stimuli. (c) SONY-FDR Camera: to record facial expression data. (d) Fingertip Pulse Oximeter (Yuwell-YX306): to measure heart rate/oxygen concentration (HR/SaO2) concurrently.

Considering the pathological subjects' limited mobility, the data collection procedure was carried out at the subject's residence, see Figure 1. The room was guaranteed to be low-noise. During the data capture, the subject wore the EGG-D100 sitting at around 3 m from the display screen, SONY-FDR camera was placed at about 50 cm away.

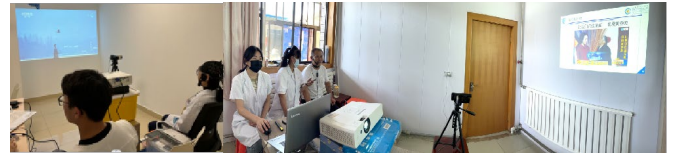

**Figure 1**: *On-site data collection.*

The integrity and intelligibility of the recordings (acoustic and glottal data) were checked by Cool Edit Pro. Considering the difficulty of recording pathological data, voice missing less than 3 words but had ample emotional expression were retention. After filtering and denoising, 4350 acoustic data and 4304 glottal data were retained, as shown in Table 2. For the convenience, naming rules oriented to materials, emotion and subject information were designed for each file in THE-POSSD database. There are 16 bits in total. Such as FS011110103012111 indicates that female patient #1 severe patient between the ages of 50 and 60 years old happily retells the third line sentence with acoustic and vocal gate data, no video data; SCL-90, dysarthria ratings, and cognitive rating scales are available.

**Table 2**: *THE-POSSD database*

| Type | Acoustic(.wav) | | Glottal (.wav) | | Video (.mp4) | HR/SaO2 (.xlsx) |
|---|---|---|---|---|---|---|
| | Emo | Non-emo | Emo | Non-emo | | |
| Qty | 2438 | 1912 | 2409 | 1895 | 43.13h | 3315 |
| Total | 4350 | | 4304 | | | |

*Qty=quantity; Emo=emotional; Non-emo=Non-emotional*

# 3. Data annotation and analysis

## 3.1. Data annotation

All future studies will be built on the accuracy of emotion expression and speech intelligibility. To fully annotate the dataset, three tasks were conducted from the aspects of emotion type, discrete dimension scoring and intelligibility rating. A WeChat mini-program was developed to carry out these jobs. For the discrete dimension scoring, valence-arousal (VA) model [24] was applied as it is quite dominated in affective computing area. Twenty college students between the age of 18 and 26 years old (10 males and 10 females) in good physical and mental health, participated in this experiment.

## 3.2. Analysis on emotion discrimination accuracy

Table 3 shows the subjective annotation accuracy of all 20 annotators on the four types of emotions. The results are promising, majorly above 87.5%, with neutral emotion having the highest accuracy. This illustrates that the emotion expression of this database is accurate and satisfies the requirements for future research.

**Table 3**: *Accuracy of emotion discrimination*

| Emotion | Neutral | Happy | Sad | Angry |
|---|---|---|---|---|
| Neutral | 90.5% | 5.4% | 5.1% | 4.2% |
| Happy | 4.5% | 87.5% | 3.3% | 2.8% |
| Sad | 2.4% | 3.3% | 87.9% | 3.6% |
| Angry | 2.6% | 3.8% | 3.6% | 89.3% |

## 3.3. Analysis of speech intelligibility

The speech intelligibility is divided into 5 levels, as seen in Table 4. From these numbers, we could discover that the disease severity is inversely proportional to the speech intelligibility. For each emotion, this trend also keeps consistent along with different severities, as exemplified in Figure 2(a). This indicates that the disease actually affects patients' communication ability and emotion expression.

**Table 4**: *Speech intelligibility of different severity*

| Intelligibility | Control | Mild | Moderate | Severity |
|---|---|---|---|---|
| Little | 0.1% | 0.4% | 4.4% | 14.6% |
| Some | 0.1% | 0.4% | 2.6% | 5.5% |
| Half | 0.5% | 1.0% | 2.6% | 4.1% |
| Mostly | 5.4% | 6.9% | 9.1% | 8.6% |
| Completely | 93.8% | 91.3% | 81.2% | 67.3% |

## 3.4. Analysis of VA spatial distribution

To assess the quality of emotion expression and their distribution in VA space, 95% confidence elliptic intervals were fitted, which makes it possible to efficiently distinguish the four emotions theoretically. As depicted in Figure 2(b), the 'happy' is separated from the valence, while the 'angry' and 'sad' are distinguished from the arousal. A small portion of the 'happy' crosses the first and second quadrants, because the subjects expressed in a more excited voice, easily confused with 'anger'. The raincloud plots (Figure 2(c)(d)) shows that all emotions obey a normal distribution with certain parameters, which means the overall concentration is more stable. One-way ANOVA results give $P_{Arousal} = 0$ and $P_{Valence} = 0$ (both $P<0.05$), indicating that there are significant differences in valence and arousal, and the four emotions are well distinguished. This paves the way for the spatial analysis of emotion recognition.

## 3.5. Correlation analysis between SCL-90 and disease severity

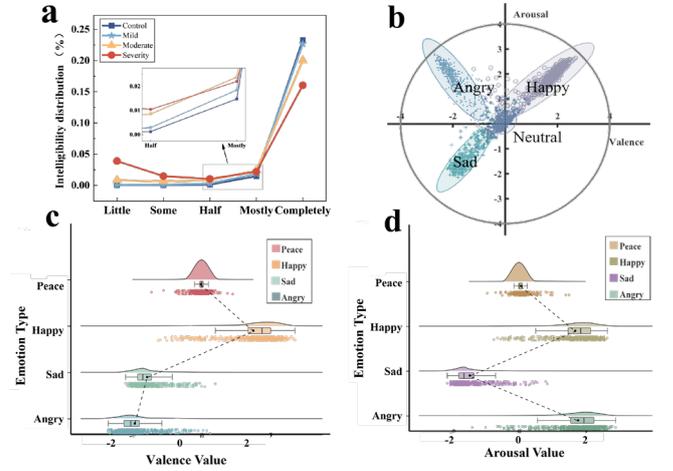

**Figure 2**: *(a)intelligibility distribution of each disease course (happy emotion), (b) VA spatial distribution, (c) valence mean, (d) arousal mean.*

The general severity index (GSI) and factor score (FS) are the evaluation indexes in SCL-90. The positive symptoms total (PST) is the number of items with a single item score ≥2. All factor > 2 are deemed "positive" by the national norm[25]. Figure 3(a) displays that the GSI, PST, and each factor is higher in the three types of disease than in controls, and the number of positive results of obsessive-compulsive disorder, interpersonal sensitivity, depression, anxiety, and psychoticism is proportional to the severity of the disease. It means that there is some sort of close association. The mean values of all the factors in the remaining pathological categories in Figure 3(b) are greater than the national norm, apart from obsessive-compulsive in mild. Furthermore, the patients in "severe" present the highest values for most of the factors. In total, the mean values exhibit an upward trend as the disease progresses. However, for normal controls, except for somatization and obsessive-compulsive, some of the remaining factors are only slightly elevated, considering the fact that the ageing of subjects and the decline of their physical functions, the data are within acceptable limits.

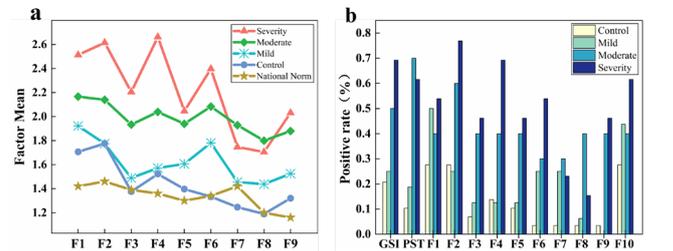

**Figure 3**: *(a) the results of positive screening check, (b) comparison of factor mean values with normal norm.*

The Kruskal-Wallis test focuses on the effects of the ten factors in SCL-90, age, gender, and education on the disease severity. According to Table 5, there is no statistically significant difference (P>0.05) between gender, education, diet and sleep (F10). Whereas, the differences in the other nine factors are significant among the different severities of the disease (P<0.05). Interpersonal sensitivity, depression, hostility, paranoia, and psychoticism are markedly different depending on the disease course (P≤0.001). It is possible to evaluate their connection to the development of the disease.

**Table 5**: *Correlation of disease with each factor, age, gender, and education*

| Items | Severe | Moderate | Mild | Control | Statistical | P |
|---|---|---|---|---|---|---|
| Age (years) | 62.0 ± 4.25 | 67.5 ± 8.88 | 71.5 ± 6.41 | 66.0±7.38 | 8.469 | 0.037** |
| Gender | 1 ± 0.48 | 1 ± 0.48 | 1 ± 0.50 | 1 ± 0.49 | 5.644 | 0.130 |
| Education | 1 ± 0.66 | 1 ± 0.52 | 1 ± 0.51 | 1 ± 0.69 | 0.032 | 0.999 |
| Somatization (F1) | 2.58 ± 0.73 | 1.88 ± 0.83 | 2.04 ± 0.63 | 1.58 ± 0.69 | 10.4 | 0.015** |
| Obsessive-Compulsive (F2) | 2.70 ± 0.89 | 2.2 ± 0.762 | 1.70 ± 0.41 | 1.70 ± 0.61 | 11.527 | 0.009*** |
| Interpersonal Sensitivity (F3) | 2.00 ± 0.87 | 1.89 ± 0.66 | 1.44 ± 0.34 | 1.22 ± 0.42 | 17.456 | 0.001*** |
| Depression (F4) | 2.92 ± 0.8 | 1.92 ± 0.76 | 1.42 ± 0.36 | 1.39 ± 0.56 | 21.747 | 0.000*** |
| Anxiety (F5) | 1.70 ± 0.87 | 1.60 ± 0.81 | 1.55 ± 0.57 | 1.30 ± 0.45 | 10.358 | 0.016** |
| Hostility (F6) | 2.70 ± 0.95 | 1.83 ± 1.03 | 1.67 ± 0.52 | 1.17 ± 0.34 | 18.291 | 0.000*** |
| Phobia (F7) | 1.57 ± 0.89 | 1.64 ± 0.92 | 1.29 ± 0.56 | 1.14 ± 0.47 | 9.084 | 0.028** |
| Paranoia (F8) | 1.50 ± 0.79 | 1.58 ± 0.54 | 1.50 ± 0.35 | 1.17 ± 0.33 | 19.678 | 0.000*** |
| Psychoticism (F9) | 2.00 ± 0.74 | 1.75 ± 0.71 | 1.55 ± 0.29 | 1.20 ± 0.38 | 16.047 | 0.001*** |
| Diet and Sleep (F10) | 2.14 ± 0.61 | 1.86 ± 0.97 | 1.93 ± 0.71 | 1.57 ± 0.65 | 6.571 | 0.087* |

*\*\*\*, \*\*, \* represent 1%, 5%, 10% significance level, respectively.*

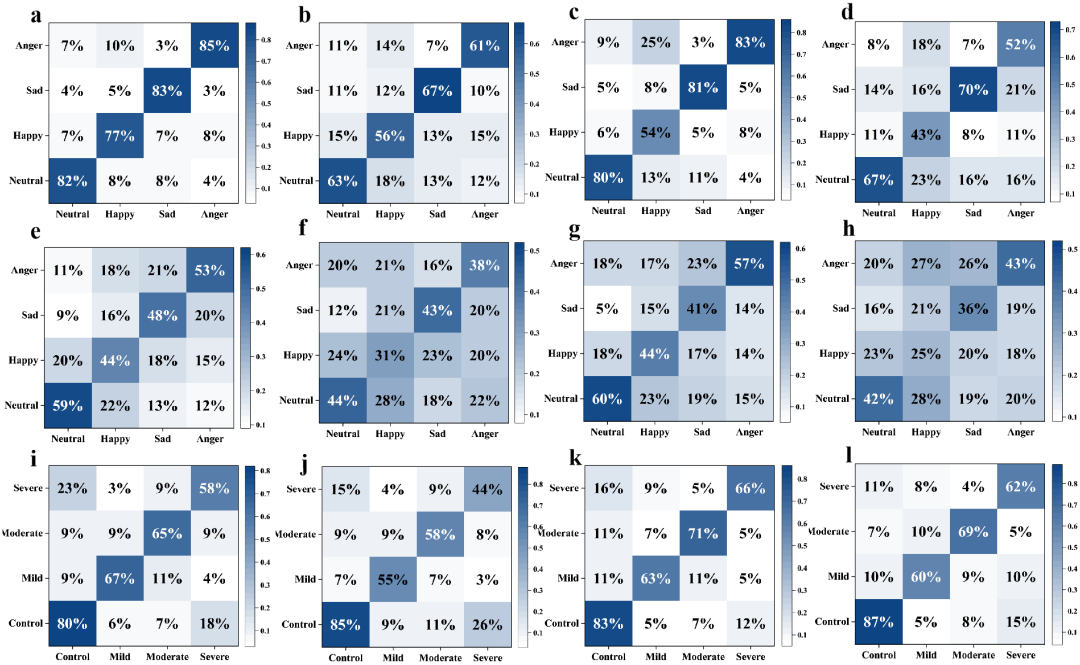

**Figure 4**: *Confusion matrix for ER (audio): (a) Control-SVM, (b) Pathological-SVM, (c) Control-RF, (d) Pathological-RF; Confusion matrix for ER (glottal): (e) Control-SVM, (f) Pathological-SVM, (g) Control-RF, (h) Pathological-RF; (i) Acoustic-Disease-SVM, (j) Acoustic-Disease-RF, (k) Glottal-Disease-SVM, (l) Glottal-Disease-RF.*

## 4. Automatic recognition test

Besides the subjective discrimination, automatic recognition test was also applied to evaluate the quality of the audio and glottal data by conducting emotion recognition (ER) and disease severity classification on support vector machine (SVM) and random forest (RF). The training and test sets were split into 7:3. The IS09 feature set (384 dimensions) was selected for emotion classification. The disease course was graded by a 101-dimensional feature set, including max, min, mean, standard deviation and variance of Pitch, PSD, LSF, MFCC(65 dimensions), and LPC (21 dimensions).

Figure 4(a-d) lists the recognition results for audio, with average accuracy of 78% for controls and 60% for patients. Happy emotion has the lowest rate and is confused with angry. High neuroticism individuals are comparatively less receptive to and less likely to keep onto the positive feelings they produce [26]. As cognitive function decline with age, high neuroticism in the elderly is more of a risk factor [27]. This result is mostly in line with the age distribution in THE-POSSD. As seen in Figure 4(e-h), the glottal data has a lower overall recognition rate than acoustic data, with average 51% for controls and 38% for patients. This is because EGG signals react differently to different subject genders and the vocal tract structures. As "sad" is difficult to express, making it tough for machines to accurately identify because the emotional features are not immediately visible. The recognition rates of the disease severities (Figure 4(i-l)) are significant, up to 87%, indicating that the pathological aspects of this database have a degree of differentiation. In conclusion, the THE-POSSD database documented successful outcomes in both emotional expression

and pathology aspect, which can serve as a reliable source of data for upcoming research needs.

## 5. Conclusions and future work

In this paper, a Chinese multimodal emotional pathological speech database (THE-POSSD) is designed and constructed for the first time. It contains 29 controls and 39 motor dysarthria patients of different severities, expressing neutral, sad, happy, anger emotions. Audio, glottal, video (Figure 5(a)), fNIRS (Figure 5(b)), heart rate/oxygen concentration data, as well as the multi-perspective information on SCL-90, MMSE, and FDA are all included. The intelligibility, emotion types, and VA model of all emotional speech were labeled, and both subjective and objective aspects of this database were examined. Besides, the SCL-90 scale analysis reveals a significant link between psychological status and the development of the disease, which is solid evidence in favor of researching psychological characteristics. The database shows a stronger clinical course difference and a more noticeable emotional separation between controls and dysarthria patients, according to all studies.

Taking the advantage of rich speech materials, THE-POSSD dataset could be applied to multiple research areas to analyse the discrimination between dysarthria patients and controls, such as with/without emotion, vowel exploration, word/sentence level analysis, fluency expression, speech intelligibility, disease severity classification, pathological speech/face recognition, emotional speech/facial expression analysis, oxyhemoglobin and deoxyhemoglobin analysis under emotional stimuli, etc. We haven't fully tested the dataset, especially for the video and fNIRS data due to limited time and experiences. Our future work will focus on these jobs. The dataset will be public on proper platforms once we finalize tests and all needed documents.

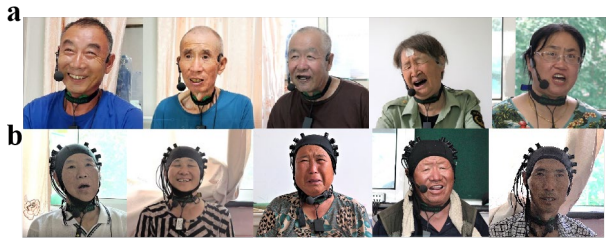

**Figure 5**: *(a)Excerpts of facial micro-expressions, (b) Near-infrared recording data capture.*

## 6. Data Availability

The data that support the findings of this study are available on request from the corresponding author, upon reasonable request. E-mail:duanshufei@tyut.edu.cn，

## 7. Acknowledgement

This study was supported by a grant from the National Natural Science Foundation of China (No. 12004275) and the China Scholarship Council. Special thanks go to all recruited participants, Taiyuan Central Hospital of Peking University First Hospital, Newcastle University for discussion and help.

## References

[1] T. Schölderle, A. Staiger, R. Lampe *et al.*, "Dysarthria in adults with cerebral palsy: Clinical presentation and impacts on communication," *Journal of Speech, Language,Hearing Research,* vol. 59, no. 2, pp. 216-229, Apr. 2016.

[2] S. Roldan-Vasco, A. Orozco-Duque, J. C. Suarez-Escudero *et al.*, "Machine learning based analysis of speech dimensions in functional oropharyngeal dysphagia," *Computer Methods Programs in Biomedicine,* vol. 208, pp. 106248-102669, Sep. 2021.

[3] M. Hireš, M. Gazda, P. Drotár *et al.*, "Convolutional neural network ensemble for Parkinson's disease detection from voice recordings," *Computers in biology medicine,* vol. 141, pp. 105021-105029, Feb. 2022.

[4] X. Menendez-Pidal, J. B. Polikoff, S. M. Peters *et al.*, "The Nemours database of dysarthric speech," in *Proceeding of Fourth International Conference on Spoken Language Processing. ICSLP'96*, Philadelphia, PA, USA, Oct. 1996, vol. 3, pp. 1962-1965: IEEE.

[5] T. A. Mesallam, M. Farahat, K. H. Malki *et al.*, "Development of the arabic voice pathology database and its evaluation by using speech features and machine learning algorithms," *Journal of healthcare engineering,* vol. 2017, pp. 878351-878351, Oct. 2017.

[6] D. Martínez, E. Lleida, A. Ortega *et al.*, "Voice Pathology Detection on the Saarbrücken Voice Database with Calibration and Fusion of Scores Using MultiFocal Toolkit," in *Advances in speech and language technologies for Iberian Languages*, Madrid, Spain, Nov. 2012, vol. 328, pp. 99-109: Springer Berlin Heidelberg.

[7] N. M. Joy and S. Umesh, "Improving acoustic models in TORGO dysarthric speech database," *IEEE Transactions on Neural Systems Rehabilitation Engineering,* vol. 26, no. 3, pp. 637-645, Mar. 2018.

[8] L. Geng, Y. Liang, H. Shan *et al.*, "Pathological Voice Detection and Classification Based on Multimodal Transmission Network," *Journal of Voice,* Dec. 2022.

[9] Z. Farhoudi and S. Setayeshi, "Fusion of deep learning features with mixture of brain emotional learning for audio-visual emotion recognition," *Speech Communication,* vol. 127, pp. 92-103, Mar. 2021.

[10] Z. Yue, E. Loweimi, Z. Cvetkovic *et al.*, "Multi-modal acoustic-articulatory feature fusion for dysarthric speech recognition," in *ICASSP 2022-2022 IEEE International Conference on Acoustics, Speech and Signal Processing (ICASSP)*, Singapore, Singapore, May. 2022, pp. 7372-7376: IEEE.

[11] L. K. Butler, S. Kiran, and H. Tager-Flusberg, "Functional near-infrared spectroscopy in the study of speech and language impairment across the life span: a systematic review," *American Journal of Speech-Language Pathology,* vol. 29, no. 3, pp. 1674-1701, Aug. 2020.

[12] C. Huo, G. Xu, W. Li *et al.*, "A review on functional near-infrared spectroscopy and application in stroke rehabilitation," *Medicine in Novel Technology Devices,* vol. 11, pp. 100064-100077, Sep. 2021.

[13] S. Liu, S. Xing, and C. Zhang, "Investigation the emotional condition of the patients with apoplex being taken bad for the

first time," *Journal of Clinical Psychosomatic Diseases,* vol. 14, no. 3, pp. 151-152, Nov. 2002.

[14] H. P. Aben, J. M. Visser-Meily, G. J. Biessels *et al.*, "High occurrence of impaired emotion recognition after ischemic stroke," *European stroke journal,* vol. 5, no. 3, pp. 262-270, Mar. 2020.

[15] C. Kim, S. Lee, I. Jin *et al.*, "Acoustic Features and Cortical Auditory Evoked Potentials according to Emotional Statues of /u/, /a/, /i/ Vowels," *Journal of Audiology Otology,* vol. 22, no. 2, pp. 80-88, Apr. 2018.

[16] L. Alhinti, H. Christensen, and S. Cunningham, "Acoustic differences in emotional speech of people with dysarthria," *Speech Communication,* vol. 126, pp. 44-60, Feb. 2021.

[17] L. Dehu, R. Qiwu, Y. Chao *et al.*, "Review on Speech Emotion Recognition Research," *Computer Engineering and Applications,* vol. 58, no. 21, pp. 40-52, Jun. 2022.

[18] CAISA Mandarin Emotional Speech Corpus, *Institute of Automation, Chinese Academy of Sciences*, 2005, http://www.chineseldc.org/resource_info.php?rid=76.

[19] C. Busso, M. Bulut, C. C. Lee *et al.*, "IEMOCAP: Interactive emotional dyadic motion capture database," *Language resources and evaluation*, vol. 42, pp.335-359, Nov. 2008.

[20] F. Burkhardt, A. Paeschke, M.,Rolfes *et al.*, "A database of German emotional speech," in *Interspeech,* Dresden, DE, Sep. 2005, vol. 5, pp. 1517-1520.

[21] L. Alhinti, S. Cunningham, and H. Christensen, "The Dysarthric Expressed Emotional Database (DEED): An audio-visual database in British English," *Plos one*, vol. 18, no. 8, pp. e0287971, Aug. 2023.

[22] L. Alhinti, H. Christensen, and S. Cunningham, "An exploratory survey questionnaire to understand what emotions are important and difficult to communicate for people with dysarthria and their methodology of communicating," *International Journal of Psychological Behavioral Sciences,* vol. 14, no. 7, pp. 187-191, Jun. 2020.

[23] Q. Du, *Mandarin Phonetics Course*. Beijing: China Radio and Television Press, 2017.

[24] R. E. Thayer, *The biopsychology of mood and arousal*. New York: Oxford University Press, 1990.

[25] W. Dang, Y. Xu, J. Ji *et al.*, "Study of the SCL-90 scale and changes in the Chinese norms," *Frontiers in Psychiatry,* vol. 11, pp. 524395-5243100, Jan. 2021.

[26] Q.-q. GE, X.-n. ZHOU, Y.-z. LIU *et al.*, "Role of neuroticism in depressive symptoms among officers and soldiers: mediating effects of negative automatic thoughts and response psychological stress," *Journal of Naval Medical University,* vol. 43, no. 7, pp. 821-826, May. 2022.

[27] M. Banjongrewadee, N. Wongpakaran, T. Wongpakaran *et al.*, "The role of perceived stress and cognitive function on the relationship between neuroticism and depression among the elderly: a structural equation model approach," *BMC psychiatry,* vol. 20, no. 1, pp. 1-8, Jan. 2020.